\documentclass[prd,a4paper,showpacs,preprint,byrevtex]{revtex4}
\usepackage{dcolumn}
\usepackage{amsmath}
\usepackage{array}
\usepackage{bm}
\usepackage{graphicx}

\begin{document}

\title{Covariant oscillator quark model for glueballs and baryons}

\author{Fabien \surname{Buisseret}}
\thanks{FNRS Research Fellow}
\email[E-mail: ]{fabien.buisseret@umh.ac.be}
\author{Claude \surname{Semay}}
\thanks{FNRS Research Associate}
\email[E-mail: ]{claude.semay@umh.ac.be}
\affiliation{Groupe de Physique Nucl\'{e}aire Th\'{e}orique,
Universit\'{e} de Mons-Hainaut,
Acad\'{e}mie universitaire Wallonie-Bruxelles,
Place du Parc 20, BE-7000 Mons, Belgium}

\date{\today}

\begin{abstract}
An analytic resolution of the covariant oscillator quark model for a
three-body system is presented. Our harmonic potential is a general
quadratic potential which can simulate both a $\Delta$-shape
configuration or a simplified Y-configuration where the junction is
located at the center of mass. The mass formulas obtained are used to
compute glueball and baryon spectra. We show that the agreement with
lattice and experimental data is correct if the Casimir scaling
hypothesis is assumed. It is also argued that our model is compatible
with pomeron and odderon approaches.
\end{abstract}



\maketitle

\section{Introduction}

The covariant oscillator quark model (COQM) is a phenomenological model
of hadrons \cite{co83}. It is not based on a Hamiltonian like usual
quark models, but on an operator giving the square mass of the
considered system. This model is not a QCD inspired one like a spinless
Salpeter Hamiltonian with a linear confining potential. But, through a
covariant formalism, it is able to deal in
a simple way with retardation effects, which are rather complicated in
other approaches \cite{sazd,olss92,Buis}. Moreover, the COQM gives the
correct Regge slopes of the mesonic and baryonic Regge
trajectories \cite{co93,co94}. An attempt to reproduce the mass spectrum
of glueballs formed of two gluons with this model has also been made in
Ref.~\cite{co97}. The key ingredient of the COQM is the particular form
of the confining potential. In a two-body system, it is a harmonic
potential depending on the separation of the particles, the spring
constant of the potential determining the value of the Regge slope. For
the
baryons, a quadratic equivalent of the so-called $\Delta$-potential has
been used in Ref.~\cite{co94}. However, recent developments in
lattice QCD rather support the picture of an Y-junction inside the
baryons \cite{Koma}, and thus the potential of the three-body COQM
should
be modified. Moreover, accurate studies have since been performed in
lattice QCD about the spectrum of glueballs \cite{lat99,lat05}. The
purpose of this paper is thus to reconsider the COQM taking into account
these new results. To do this, we will use a more physically relevant
three-body potential, simulating an Y-junction, and we will show that
the mass
spectra and the wave functions can be analytically found. This is
somewhat unusual in a three-body problem. Solving the equations of the
COQM will give us the possibility to compare its predictions with the
well known baryon spectra, and new lattice QCD results concerning the
glueballs. Moreover, it will give us a first estimation of the
retardation effects in a three hadrons system, which we hope will open
the way for a three body generalization of the work in Ref.~\cite{Buis}.

\par Our paper is organized as follows. In Sec.~\ref{mesonic}, we
introduce the formalism of the COQM through the simple case of a
two-body system, as it is done in Ref.~\cite{co93}. Then, in
Sec.~\ref{baryonic}, we discuss the introduction of a general potential
for the three-body COQM, which can be seen as a mixing of a harmonic
equivalent of the Y-junction and the harmonic potential of
Ref.~\cite{co94}. After that, we show that it is possible to find
analytical solutions for the three-body COQM with our new potential. In
Sec.~\ref{compar}, we discuss the value of the parameters of our model
and we compare its results with the experimental and lattice QCD data.
Finally, we sum up our results in Sec.~\ref{conclu}.

\section{Two-body COQM}\label{mesonic}

The equation of the COQM for $N$ interacting hadrons is \cite{co83}
\begin{equation}\label{coeq1}
\sum^{N}_{k=1}\frac{\bm{p}^{2}_{k}}{2m_{k}}+U(\bm{x}_{i}-\bm{
x}_{j})=0,
\end{equation}
where the potential $U$, assumed to depend only on relative
coordinates, describes the confining interaction between $N$
particles whose four-vector coordinates are $\bm{x}_{k}$.
We can perform a change of coordinates and express the
$\bm{x}_{i}$ in terms of the center of mass position, $\bm{R}$, and
$N-1$ relative coordinates, $\bm{r}_{a}$. Then, Eq.~(\ref{coeq1}) reads
\begin{equation}\label{coeq2}
\frac{\bm{P}^{2}}{2m_{t}}+\sum^{N-1}_{a=1}\frac{\bm{\pi}^{2}_{a}}{2\mu_{
a}}+U(\bm{r}_{b})=0,
\end{equation}
where $\bm{P}$ and $\bm{\pi}_{a}$ are the momenta associated to $\bm{R}$
and $\bm{r}_{a}$ respectively. The $\mu_{a}$ are the reduced masses and
$m_{t}=m_{1}+\dots+m_{n}$.
But, we know that $\bm{P}^{2}=-M^{2}$, where $M$ is the mass of the
system. So, instead of a Hamiltonian, the COQM allows us to write an
equation giving the square mass of the system from Eq.~(\ref{coeq2})
\begin{equation}\label{coeq3}
M^{2}=2m_{t}  \left[\sum^{N-1}_{a=1}\frac{\bm{\pi}^{2}_{a}}{2\mu_{a}}+U(
\bm{r}_{b})\right].
\end{equation}

In this section, we will only consider the quark-antiquark system. The
confining potential for mesons is \cite{co93}
\begin{equation}
U=\frac{1}{2}K_{2} \bm{r}^{2}-W,
\end{equation}
with $\bm{r}=\bm{x}_{1}-\bm{x}_{2}\equiv (\sigma,r,\theta,\varphi)$. The
spring constant $K_{2}$ is a parameter of the model, as well as the
constant $W$, which can take into
account in a very simple way other contributions than the confinement:
One gluon exchange,
spin interactions,\dots \cite{sem}. The quantized version of
Eq.~(\ref{coeq3}) can be written in this case as
\begin{equation}\label{meson}
2m_{t}\left[\frac{\bm{p}^{2}}{2\mu}+\frac{1}{2}K_{2} \bm{r}^{2}-W\right]
\left|\psi\right\rangle=M^{2}\left|\psi\right\rangle,
\end{equation}
with $\mu=m_{1}m_{2}/m_{t}$ the reduced mass. Applying the well known
theory of the harmonic oscillator, one finds
\begin{equation}\label{coqm_m}
M^{2}=2m_{t}\sqrt{K_{2}/\mu}\ (2n+\ell+1)-2m_{t}W,
\end{equation}
\begin{equation}\label{phico}
\psi=\chi_{0}(\sigma)\phi_{n,\ell}(r)Y^{m}_{\ell}(
\theta,\varphi),
\end{equation}
where the $Y^{m}_{\ell}$ are the spherical harmonics. With
$\beta=\sqrt{\mu K_{2}}$,
\begin{equation}\label{osc1d}
\chi_{n}(x)=\left(\frac{\beta}{\pi}\right)^{1/4}\frac{1}{\sqrt{2^{n}n!}}
\ e^{-\beta x^{2}/2}\, H_{n}(\beta x)
\end{equation}
is an eigenfunction of the one dimensional harmonic oscillator, and
\begin{equation}\label{osc3d}
\phi_{n,\ell}=\beta^{\frac{1}{2}(\ell+\frac{3}{2})}\sqrt{\frac{2n!}{
\Gamma(n+\ell+\frac{3}{2})}}\ r^{\ell}\, e^{-\beta r^{2}/2}\,  L^{\ell+
\frac{1}{2}}_{n}(\beta r^{2})
\end{equation}
is a radial eigenfunction of the three dimensional
harmonic oscillator \cite{Brau}.
$H_{n}$ and $L^{\alpha}_{n}$ are the Hermite and Laguerre polynomials
respectively.
It is worth mentioning that the usual factor $(2n+\ell+3/2)$ of the
harmonic oscillator is here replaced by $(2n+\ell+1)$ in
Eq.~(\ref{coqm_m}). To understand this, we have to consider the
contribution of
the relative time $\sigma$. When Eq.~(\ref{meson}) is solved, two
harmonic oscillators appear: one for the spatial part and one for the
relative time part. The second has the opposite sign of the first and
will contribute to decrease the square mass. But a nonphysical degree of
freedom is now present: which eigenstate of the relative time oscillator
do we have to choose? The prescription, which can be written in
a covariant way, is to consider only the
fundamental state as a physical one \cite{sazd,co93}. That is why we
find $\chi_{0}$ in relation~(\ref{phico}) and $-m_{t}\sqrt{K_{2}/\mu}$
for the
contribution of the relative time. This contribution is in fact due to
the retardation effects as they appear in the COQM. The most probable
value for $\sigma$ is $0$, which is in
agreement with the hypothesis of Ref.~\cite{Buis}.
When the two particles have the same mass $m$, formula~(\ref{coqm_m})
reduces to
\begin{equation}\label{coqm_m2}
M^{2}=\sqrt{32 m K_{2}}\ (2n+\ell+1)-2 m W.
\end{equation}

It has been shown in Ref.~\cite{co93} that the COQM is able to
predict the meson Regge
trajectories in agreement with the experimental data with $K_{2}=0.107$
GeV$^{3}$. We see
from Eq.~(\ref{coqm_m}) that the COQM should not be used for heavy
mesons, because only light mesons exhibit Regge trajectories. Finally,
we can remark that the COQM is not relevant for massless particles. This
drawback is inherent to this approach, but it is not troublesome if
the constituent quark and gluon masses are used instead of the current
ones.

\section{Three-body COQM}\label{baryonic}

\subsection{The confining potential}

As in the previous section, we will only consider quark systems, that is
to say baryons. We summarize here some considerations of
Ref.~\cite{co94}. The potential
which is used in this paper has the form
\begin{equation}
U=\frac{1}{2} K_{3} \sum_{i<j}\left(\bm{x}_{i}-\bm{x}_{j}\right)^{2}.
\end{equation}
It can be seen as an harmonic equivalent of the usual
$\Delta$-potential. With the same quark masses as in the mesonic case, a
good agreement between
theoretical and experimental baryonic Regge slopes can be obtained if
$K_{3}=K_{2}/4$ \cite{co94}. The proposed justification of such a $1/4$
factor is
the following heuristic argument: in a usual meson model with linear
confinement, the potential term of the Hamiltonian reads $a_{2}r$, where
$a_{2}$ is the energy density of the flux tube between the quark and the
antiquark and $r$ is their spatial separation. In the COQM, the
potential appearing in the square mass operator is $K_{2}r^{2}$.
So, one can consider that $K_{2}\propto a_{2}^{2}$. Then, if we assume
that the energy density $a_{3}$ of the flux tube between two quarks $i$
and $j$ in a baryon is proportional to the color Casimir operator
$\tilde{\lambda}_{i}\tilde{\lambda}_{j}/4$, we should have
$a_{3}=a_{2}/2$ and thus $K_{3}=K_{2}/4$ for a $\Delta$-shape
potential.

However, recent developments in lattice QCD tend to confirm the
Y-junction as the more realistic configuration for the flux tube in the
baryons \cite{Koma}. In this picture, flux tubes start from each quark
and meet at the Toricelli point of the triangle formed by the three
particles. This point, denoted $\vec x_{T}$, is such that it minimizes
the sum of the flux tube lengths, and its position is a complicated
function of the quark coordinates $\vec{x}_{i}$. Moreover, the energy
density of the tubes appears to be equal for mesons
and hadrons:
$a_{3}=a_{2}$. As we want to include these
recent results in the COQM, we have to change the expression of the
harmonic potential. In particular, we have to keep $K_{3}=K_{2}$ and to
take a quadratic equivalent of the Y-junction.

In Ref.~\cite{Bsb04}, the complex Y-junction potential
\begin{equation}
V=a_{3} \sum^{3}_{i=1} \left|\vec{x}_{i}-\vec{x}_{T}\right|
\end{equation}
is approximated by the more easily computable expression
\begin{equation}\label{pot1}
V=a_{3}\left[\alpha \sum^{3}_{i=1}\left|\vec{x}_{i}-\vec{R}\right|  +(1-
\alpha)\frac{1}{2}\sum_{i<j}\left|\vec{x}_{i}-\vec{x}_{j}\right|\right],
\end{equation}
where $\vec{R}$ is the position of the center of mass. If $\alpha=1$,
Eq.~(\ref{pot1}) is a simplified Y-junction, where the Toricelli point
is replaced by the center of mass. If $\alpha=0$, this interaction
reduces
to a $\Delta$-potential. Let us note that the presence of the factor
$1/2$ in the $\Delta$-part of the potential is purely geometrical and
simply arises because in a triangle $ABC$ with a Toricelli point $T$,
$\left|AT\right|+\left|BT\right|+\left|CT\right|\geq(\left|AB\right|+
\left|BC\right|+\left|CA\right|)/2$. Results of
Ref.~\cite{Bsb04}, obtained in the framework of a potential model, show
that $\alpha=1$ gives better results than $\alpha=0$, and that the
Y-junction is approximated at best when $\alpha$ is close to 1/2.

In order to simulate at best the genuine Y-junction, keeping the
calculations feasible, we assume, in agreement with  Eq.~(\ref{pot1}),
the following expression for the potential in the baryonic COQM
\begin{equation}\label{pot2}
U=\frac{1}{2}K_{3}\left[\alpha \sum^{3}_{i=1}\left(\bm{x}_{i}-\bm{R}
\right)^{2} +(1-\alpha)\frac{1}{4}\sum_{i<j}\left(\bm{x}_{i}-\bm{x}_{j}
\right)^{2}\right]-W,
\end{equation}
with $K_{3}=K_{2}$. The origin of the factor $1/4=(1/2)^{2}$ is now seen
as geometrical only. We define
\begin{equation}
\mu=\alpha K_{3},\ \rho=K_{3}(1-\alpha)/4,
\end{equation}
and the potential (\ref{pot2}) becomes
\begin{equation}\label{pot3}
U=\frac{\mu}{2} \sum^{3}_{i=1}\left(\bm{x}_{i}-\bm{R}\right)^{2}+\frac{
\rho}{2} \sum_{i<j}\left(\bm{x}_{i}-\bm{x}_{j}\right)^{2}-W,
\end{equation}
which is the expression we use in the following.

\subsection{Mass formula and wave function}

We have to solve Eq.~(\ref{coeq1}), which in this case reads
\begin{equation}\label{res1}
\sum^{3}_{i=1}\frac{\bm{p}^{\ 2}_{i}}{2m_{i}}+U=0,
\end{equation}
with $U$ given by Eq.~(\ref{pot3}). First of all, we will replace the
quark coordinates
$\bm{x}_{i}=\left\{\bm{x}_{1},\bm{x}_{2},\bm{x}_{3}\right\}$ by
$\bm{x}'_{k}=\left\{\bm{R},\bm{\xi},\bm{\eta}\right\}$, with the center
of mass defined as
\begin{equation}\label{cmdef}
\bm{R}=\frac{m_{1}\bm{x}_{1}+m_{2}\bm{x}_{2}+m_{3}\bm{x}_{3}}{m_{t}}.
\end{equation}
$m_{t}=m_{1}+m_{2}+m_{3}$ and $\{\bm{\xi},\bm{\eta}\}$ are two relative
coordinates. The change of coordinates is made via a matrix $Q$, thanks
to the relation $\bm{x}_{i}=Q_{ik}\bm{x}'_{k}$. Let us note that the
invariance of the Poisson brackets demands that
$\bm{p}_{i}=\left(Q^{-1}\right)^{\rm{T}}_{ik}\bm{p}'_{k}$, with
$\bm{p}'_{i}=\left\{\bm{P},\bm{\pi}_{\xi},\bm{\pi}_{\eta}\right\}$. We
define
\begin{equation}
Q=\left( \begin{array}{ccc}
 1 & A & B\\
 1 & C & D\\
 1 & E & F
 \end{array}\right),
\end{equation}
and find that the elements of $Q$ can be constrained by the following
equations
\begin{eqnarray}
A&=&-\frac{m_{2}}{m_{1}}C-\frac{m_{3}}{m_{1}}E,\label{eq1}\\
B&=&-\frac{m_{2}}{m_{1}}D-\frac{m_{3}}{m_{1}}F,\label{eq2}\\
C&=&\frac{1}{m_{t}F}\left(m_{1}+m_{t}ED\right),\label{eq3}\\
D&=&F \delta,\label{eq4}\\
E&=&-\frac{m_{1}m_{2}}{m_{t}}\frac{\left[D(m_{1}+m_{2})+F m_{3}\right]}{
\left[\left(D m_{2}+F m_{3}\right)^{2}+D^{2}m_{1}m_{2}+F^{2}m_{1}m_{3}
\right]}\label{eq5},
\end{eqnarray}
where $\delta$ is a solution of
\begin{eqnarray}
\label{eq6}
&&m_{1}-m_{2})m_{2}(m_{t}\rho+m_{3}\mu)\delta^{2}+(m_{2}-m_{3
}) \left[ m_{2}m_{3}\mu-m_{1}m_{t}(\mu+2\rho)\right]\delta \nonumber
\\
&&+(m_{3}-m_{1})m_{3}(m_{t}\rho+m_{2}\mu)=0.
\end{eqnarray}
Constraints (\ref{eq1}) and (\ref{eq2}) are consequences of the
definition (\ref{cmdef}). Equation (\ref{eq3}) ensures that $\det Q=1$
in order to simplify the calculations of the $\bm{p}'_{i}$. These three
relations are sufficient to cancel the terms containing the cross
products $\bm{P}\cdot\bm{\pi}_{\xi}$ and $\bm{P}\cdot\bm{\pi}_{\eta}$ in
the kinetic part of relation (\ref{res1}). The last cross product
$\bm{\pi}_{\xi}\cdot\bm{\pi}_{\eta}$ vanishes thanks to Eq.~(\ref{eq5}).
Finally, formulas~(\ref{eq4}) and (\ref{eq6}) suppress the terms
containing the cross product $\bm{\xi}\cdot\bm{\eta}$ in the
potential~(\ref{pot3}).

The last parameter to fix is $F$, which has to be nonzero. We define
\begin{eqnarray}
\Gamma&=&\left[(\delta m_{2}+m_{3})^{2}+m_{1}(\delta^{2}m_{2}+m_{3})
\right]/m_{1},\\
\phi&=&\sqrt{\frac{m_{1}m_{2}m_{3}}{m_{t}}},
\end{eqnarray}
and we choose
\begin{equation}
F^{2}=\phi/\Gamma.
\end{equation}
We can then rewrite Eq.~(\ref{res1}) as
\begin{equation}\label{res2}
\frac{\bm{P}^{2}}{2m_{t}}+\frac{\bm{\pi}^{2}_{\xi}}{2\phi}+
\frac{\bm{\pi}^{2}_{\eta}}{2\phi} +\frac{1}{2}\Omega_{\xi}\bm{\xi}^{2}+
\frac{1}{2}\Omega_{\eta}\bm{\eta}^{2}-W=0,
\end{equation}
where
\begin{equation}
\Omega_{\xi}=\mu(A^{2}+C^{2}+E^{2})+\rho\left[(A-C)^{2}+(A-E)^{2}+(C-E)^
{2}\right],
\end{equation}
 \begin{equation}
\Omega_{\eta}=\mu(B^{2}+D^{2}+F^{2})+\rho\left[(B-D)^{2}+(B-F)^{2}+(D-F)
^{2}\right].
\end{equation}

Equation~(\ref{res2}) has the nice property that the variables are
all separated. Using the same arguments as those discussed in
Sec.~\ref{mesonic}, the square mass spectrum can be
analytically computed. It reads
\begin{equation}\label{barform}
M^{2}=2m_{t}\left[(2n_{\xi}+\ell_{\xi}+1)\omega_{\xi}+(2n_{\eta}+\ell_{
\eta}+1)\omega_{\eta}-W\right],
\end{equation}
with
\begin{equation}
\omega_{\xi}=\sqrt{\Omega_{\xi}/\phi},\ \ \beta_{\xi}=\phi\omega_{\xi},\
\ \omega_{\eta}=\sqrt{\Omega_{\eta}/\phi},\ \ \beta_{\eta}=\phi\omega_{
\eta}.
\end{equation}
The internal wave function is given by
\begin{equation}\label{fo}
\psi=\chi_{0}(\xi_{0})\phi_{n_{\xi},\ell_{\xi}}(\xi)Y^{
m_{\xi}}_{\ell_{\xi}}(\theta_{\xi},\varphi_{\xi})\chi_{0}(\eta_{0})\phi_
{n_{\eta},\ell_{\eta}}(\eta)Y^{m_{\eta}}_{\ell_{\eta}}(\theta_{\eta},
\varphi_{\eta}),
\end{equation}
where we used the definitions (\ref{osc1d}) and (\ref{osc3d}). Only the
ground state of the temporal part of the wave function must be
considered, following the prescription of Ref.~\cite{co93}. It is worth
mentioning that the most probable values for the relative times are
$\xi^{0}=\eta^{0}=0$, like in the mesons. It can
be observed from Eq.~(\ref{barform}) that the retardation effects bring
a negative contribution to the squared mass, given by
$-m_t (\omega_\xi+\omega_\eta)$.
\par As the COQM is expected only to be valid for light particles, the
four possible baryonic systems are: $nnn$, $snn$, $ssn$ or $sss$ ($n$
stands for $u$ or $d$). Thus,
we can always consider simplified cases where at least two masses are
equal. When $m_{2}=m_{3}=m$, a solution of Eq.~(\ref{eq6}) is
$\delta=-1$, and one can find that
\begin{equation}
\omega_{\xi}=\sqrt{\frac{(m_{1}+2m)^{2}\rho+(m^{2}_{1}+2m^{2})\mu}{m_{1}
m\, m_{t}}},
\end{equation}
\begin{equation}
\omega_{\eta}=\sqrt{\frac{\mu+3\rho}{m}}.
\end{equation}
When the three masses are equal, we have simply
\begin{equation}
\omega_{\xi}=\omega_{\eta}=\sqrt{\frac{\mu+3\rho}{m}}=\omega,
\end{equation}
and the square mass formula reduces to
\begin{equation}\label{mass3m}
M^{2}=3\sqrt{ (3+\alpha)K_{3}m}\ \left[2(n_{\xi}+n_{\zeta})+(\ell_{\xi}+
\ell_{\eta})+2\right]-6 m W.
\end{equation}
It can be checked that for the particular case $\alpha=0$, our solution
agrees with the one of Ref.~\cite{co94}. For three equal masses, a
variation of $\alpha$ can be simply simulated by a variation of $K_3$;
this is not the case when particles have different masses.

\section{Comparison with experimental and lattice QCD data}
\label{compar}

Since the COQM includes neither the spin ($S$) nor the isospin
($I$) of the constituent particles, the data which can be
reproduced here are the spin and isospin average masses, denoted
$M_{\rm av}$. These are given by the following relation \cite{Brau98}
\begin{equation}\label{mav}
M_{{\rm av}}=\frac{\sum_{I,J}(2I+1)(2J+1) M_{I,J}}{\sum_{I,J}(2I+1)(2J+
1)},
\end{equation}
with $\vec J=\vec L+\vec S$, and where $M_{I,J}$ are the different
masses of
the states with the same orbital angular momentum $\ell$ and the same
quark content. We also use the formula~(\ref{mav}) to compute a mass
with the three-body COQM, but in this case $I$ and $J$ are replaced by
the values of $\ell_{\xi}$ and $\ell_{\eta}$ corresponding to a given
$\ell=\ell_{\xi}+\ell_{\eta}$.

\subsection{Mesons and baryons}
\label{mesbar}

\par It is a well known fact that the Regge slope of light mesons,
such as $n\bar{n}$ states, is roughly equal to the Regge slope of the
corresponding baryons $nnn$ \cite{Fabre}. It is readily seen that the
slope of the mass formula for meson~(\ref{coqm_m2}) and the slope of the
mass formula for baryon~(\ref{mass3m}) are equal if
\begin{equation}
\alpha=\frac{5}{9}\approx 0.56,
\end{equation}
a value close to the optimal one of $0.5$ found in Ref.~\cite{Bsb04}. In
the following, this value of $\alpha$ will be always used.

\begin{figure}[ht]
\includegraphics*[height=10cm]{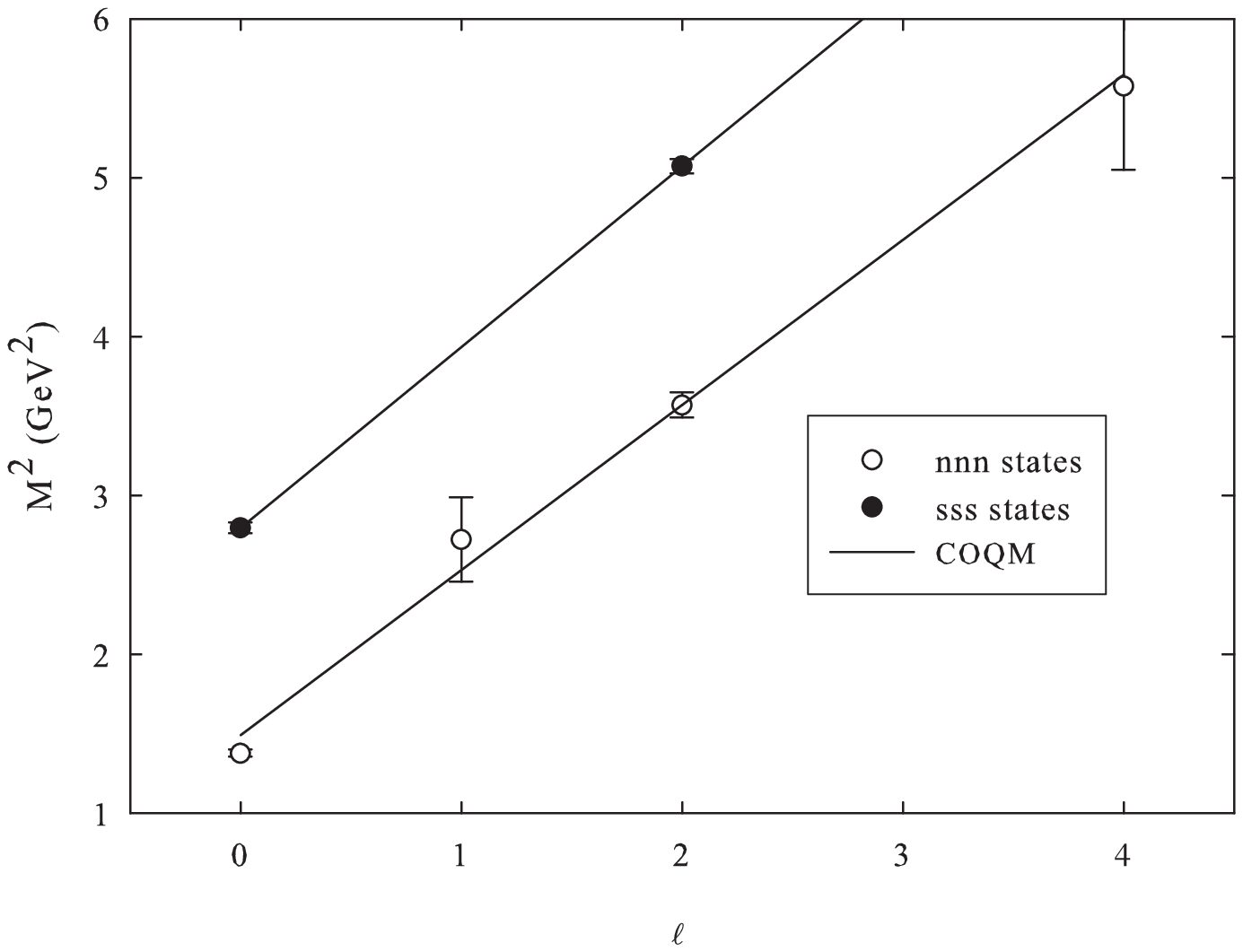}
\caption{Average experimental square masses of the $nnn$ states
(empty circles) and $sss$ states (full circles), compared with the
theoretical results (straight lines). The $nnn$ states are
computed from the $N$ and $\Delta$ baryons, and
the $sss$ states are the $\Omega$ baryons ($W_{nnn}=0.312$ GeV and
$W_{sss}=-0.232$ GeV).}
\label{fig:coqm}
\end{figure}

\begin{figure}[h]
\includegraphics*[height=10cm]{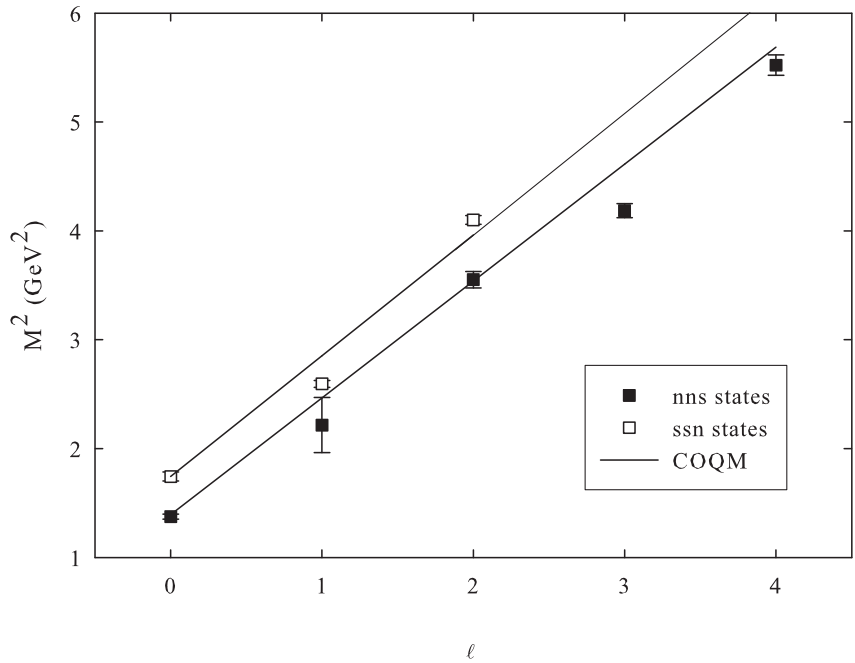}
\caption{Same as Fig. \ref{fig:coqm}, but for the $nns$ (empty squares)
and $ssn$ states (full squares). The $nns$ states are computed from
the $\Lambda$ and $\Sigma$ baryons, and the $ssn$ are
the $\Xi$ baryons ($W_{nns}=0.379$ GeV and $W_{ssn}=0.223$ GeV).}
\label{fig:coqm2}
\end{figure}

Like in all potential models, the strength of the confining potential
and the constituent quark masses must be fixed on data.
We take the spring constants $K_2=K_3$, as it is argued in
Sec.~\ref{baryonic}, and we fix the value of this parameter at
$0.107$~GeV$^3$ as in Ref.~\cite{co93}. The
constituent quark masses are chosen to be equal to $m_{n}=0.313$ GeV and
$m_{s}=0.375$ GeV, in order to obtain good baryon spectra (these values
are different from those used in Ref.~\cite{co93}). The results
are plotted in Figs.~\ref{fig:coqm} and \ref{fig:coqm2}.
We can observe a good agreement between the COQM and the average
square masses. What is important is that the correct slope is obtained,
since the absolute values of the masses depend on ad hoc values of the
parameter $W$.
$W$ is positive in every case, excepted for the $sss$ baryons. This
could be caused by strong spin interactions: masses of the $sss$
states considered are not average ones. Our data concerning the baryons
are taken from Ref.~\cite{pdg}.

The difference between the constituent masses $m_n$ and $m_s$ can seem
too small, but such a mass difference is obtained in potential models in
which a constituent state dependent mass is defined as
$\langle \sqrt{\vec p \,^2 + \nu^2} \rangle$ where $\nu$ is the current
mass. The value of $m_s-m_n$ is around 40~MeV in
Refs.~\cite{naro04,sema04} for ground states of mesons and baryons. In
Ref.~\cite{kerb00}, an analytical approximate formula is given for the
constituent quark mass of the baryon ground states : a difference of
about 30~MeV is found for current masses $\nu_n=0$ and  $\nu_s=130$~MeV
\cite{pdg}. In our model, the constituent masses are not state-dependent
but, in these references, small values for the difference $m_s-m_n$ are
also found, and absolute masses $m_n$ and $m_s$ are in agreement with
our values.

Formula~(\ref{coqm_m}) implies that the square meson masses $M^2$ have a
linear dependence on the orbital angular momentum $\ell$
\begin{equation}
\label{m2beta}
M^2=\beta \ell +\beta_0.
\end{equation}
This relation is also well verified experimentally. Using the data and
the procedure of Ref.~\cite{Buis05}, average square meson masses can be
computed, and the relation~(\ref{m2beta}) can be fitted on these data to
obtain an average experimental Regge slope $\beta_{\text{exp.}}$. With
our parameters optimized for baryon spectra, a theoretical Regge slope
$\beta_{\text{COQM}}$ can be computed for mesons. It can be seen on
Table~\ref{tab:meson} that the two slopes differ only by around 6\% for
various $q\bar q$ systems.
By increasing the value of $m_n$ by about 30~MeV, it is possible to
improve the theoretical meson spectra. The price to pay is a slight
deterioration of the baryon spectra.

\subsection{Glueballs}
\label{glueb}

We will assume here that the energy density of the flux tube starting
from a particle $i$ is proportional to the Casimir operator
$\sum_b T^{(i)}_b T^{(i)}_b$, with $T^{(i)}_b$ being a $SU(3)$ generator
in the corresponding representation. Several approaches tend to confirm
this hypothesis \cite{scaling,new}. Then, if $a_2$ is the energy density
in a meson, we should have $a_{2g}=(9/4)a_2$ in a glueball formed of two
gluons, and $a_{3g}=a_{2g}$ where $a_{3g}$ is the energy density in a
three-gluon glueball. In the COQM, the corresponding spring constants
$K_{3g}$ and
$K_{2g}$ will thus be given by $K_{3g}=K_{2g}=(9/4)^2 K_2$, where $K_2$
is the spring constant for a meson. In order to simulate at best the
Y-junction we will also use $\alpha=5/9$ as in the baryonic sector.

The constituent gluon mass $m_g$ is a parameter of the model which must
be fixed as the constituent quark masses.  Because of the scarcity of
reliable experimental data, we will determine it by using lattice-QCD
results about two-gluon glueball spectrum.. We consider that all the
positive charge conjugate states given in Refs.~\cite{lat99,lat05} are
two-gluon glueballs. This in agreement with the results of the potential
model of Ref.~\cite{Brau04}. These states are shown in
Table~\ref{tab:glue} with the average square masses computed using
Eq.~(\ref{mav}). By fitting the mass formula~(\ref{coqm_m2}) on these
data, we obtain $m_g=0.770\pm 0.340$~GeV, a value close to the usual
ones \cite{Brau04,new,abre05}. Formula~(\ref{coqm_m2}) shows that the
two-gluon  glueball mass depends on the product $K_{2g} m_g$. If we
assume a color-dependence on the square-root of the Casimir operator
\cite{john76}, it is necessary to use a gluon mass around 1.7~GeV,
which seems too heavy.

According to the glueball-pomeron theory, the $J^{++}$ two-gluon
glueballs, having
a maximum intrinsic spin $S$ coupled to a minimum possible orbital
angular momentum $\ell$, stand on Regge trajectories. In
Ref.~\cite{donn98}, the following relation is obtained
\begin{equation}
\label{rtrajgg1}
J=0.25\,M^2_{gg} + 1.08.
\end{equation}
This result is close to the one obtained in Ref.~\cite{meye05}.
In Ref.~\cite{new}, the relation obtained is noticeably different
\begin{equation}
\label{rtrajgg2}
J=0.36\,M^2_{gg} + {0.80\atop 0.57}.
\end{equation}
The value of the ordinate at origin in this model depends on the value
chosen for the strong coupling constant.
If we use our average square two-gluon glueball masses (see
Table~\ref{tab:glue}) and if we assume $J=\ell+2$ (maximum
$J$-coupling), we find
\begin{equation}
\label{rtrajgg3}
J=(0.27\pm 0.06)\,M^2_{gg} + (0.59\pm 0.49).
\end{equation}
One can see that our slope is in agreement with of the slope found in
Ref.~\cite{donn98}, but our ordinate at origin is more compatible with
the corresponding one in Ref.~\cite{new}. Finally,
Relation~(\ref{rtrajgg3}) is in good agreement with the relation found
in the COQM of Ref.~\cite{co97}
\begin{equation}
\label{rtrajgg4}
J=0.26\,M^2_{gg} + 0.67.
\end{equation}

If the interaction between gluons was spin-independent, we could expect
that the lightest three-gluon glueballs are the states with a vanishing
total orbital angular momentum and $J^{PC}=0^{-+}$, $1^{--}$,
$3^{--}$ \cite{llan05}. The lattice calculations yield very different
results \cite{lat99,lat05}. This can be due to very large spin-orbit
effects, which have been observed in two-gluon glueballs \cite{Brau04}.
In this situation, we can expect that our spinless COQM can just
describe the three-gluon glueballs with a vanishing
total orbital angular momentum. An average square mass is obtained in
Table~\ref{tab:glue} with the $1^{--}$, $3^{--}$ states of
Ref.~\cite{lat05}. The $0^{-+}$ state of this work is assumed to be a
two-gluon glueball (see above).

The equivalent of the pomeron for three-gluon glueballs is called the
odderon. $J^{--}$ three-gluon glueballs, having
a maximum intrinsic spin coupled to a minimum possible orbital
angular momentum, are also expected to stand on Regge
trajectories.
The following results are predicted in Ref.~\cite{llan05} for a QCD
effective Hamiltonian and for a non relativistic potential model
respectively
\begin{eqnarray}
\label{rtrajggg}
J_{\text{eff.}}&=&0.23\,M^2_{ggg} - 0.88, \\
J_{\text{NR}}&=&0.18\,M^2_{ggg} + 0.25.
\end{eqnarray}
Let us note that a slope for $ggg$ states which is half the slope for
$gg$ states has been suggested \cite{meye05}.

If we fit our square mass formula~(\ref{mass3m}) with the unique data
of Table~\ref{tab:glue} for three-gluon glueballs, and if we assume
$J=\ell+3$ (maximum $J$-coupling) we obtain
\begin{equation}
\label{rtrajggg2}
J=(0.27\pm 0.06)\,M^2_{ggg} - (1.57\pm 1.54).
\end{equation}
The same slope is predicted for
two-gluon and three-gluon glueballs, as in the case of quark systems,
because we assume $\alpha=5/9$ and $K_{3g}=K_{2g}$, and also because we
choose for the gluon mass the value of 0.770~GeV found for the two-gluon
glueballs. Our result is then compatible with the one from the QCD
effective Hamiltonian of Ref.~\cite{llan05}.

It is worth mentioning that the Regge trajectories presented in
Refs.~\cite{donn98,new,llan05,meye05} concern glueballs with given
$J^{PC}$.
Our model can only predict spin average square masses. So, the absolute
masses given by our model cannot, in principle, be compared directly
with masses of states with given $J^{PC}$. The ordinates at origin of
Regge trajectories calculated with our model could then have a
significant error, but we can expect a correct computation of the slope.

\section{Summary of the results}\label{conclu}

In this paper, we have solved the covariant oscillator quark model
applied to three-body systems, with a general quadratic potential. This
interaction is a linear combination between a $\Delta$-potential and a
simplified Y-junction where the Toricelli point is identified with the
center of mass; it is believed to be a good approximation of the true
Y-junction potential. We found analytic expressions for the mass
spectrum and for the total wave function. We then have shown that our
results are in quite good agreement with experimental and lattice data
concerning baryons and glueballs provided the Casimir scaling is true.
However, a more detailed study of the glueballs with the COQM requires
to take into account the spin interactions, which are particularly
strong in these particles.

\acknowledgments

The authors would thank the FNRS Belgium for financial support.

\clearpage

\clearpage

\begin{table}[h]
\caption{Regge slope $\beta$ for mesons (see formula~(\ref{m2beta})).
$\beta_{\text{exp.}}$ is the value obtained by averaging experimental
data (see Ref.~\cite{Buis05}) and $\beta_{\text{COQM}}$ is our
theoretical value (see formula~(\ref{coqm_m})). The values of parameters
$W$ for the various systems are also given.
\label{tab:meson}}
\begin{ruledtabular}
\begin{tabular}{ccccc}
State & $\beta_{\text{exp.}}$ (GeV$^{-2}$) &
$\beta_{\text{COQM}}$ (GeV$^{-2}$) & $W$ (GeV) \\
\hline
$n \bar n$ & 1.13 & 1.04 & 0.370 \\
$n \bar s$ & 1.16 & 1.09 & 0.220 \\
$s \bar s$ & 1.19 & 1.13 & 0.065 \\
\end{tabular}
\end{ruledtabular}
\end{table}

\begin{table}[h]
\caption{Average square masses, $M^{2}_{\text{av}}$, for glueball
states. The third and fourth columns show the different states used to
compute the quantities $M^{2}_{\text{av}}$. Data are taken from
Ref.~\cite{lat05}, except the mass of the $0^*$ state which is taken
from Ref.~\cite{lat99}.
\label{tab:glue}}
\begin{ruledtabular}
\begin{tabular}{ccccc}
State\  & $(n+1)L$    & $J^{PC}$ & $M$ (GeV)& $M^{2}_{\rm{av}}$(GeV$^2$)
\\ \hline
$gg$& 1S\ \ \ & \ \ 0$^{++}$\ \  & $1.710\pm 0.130$\ \  &
$5.183\pm 0.667$ \\
 & & \ \ 2$^{++}$\ \ & $2.390\pm 0.150$\ \  &   \\
&1P\ \ \ & \ \ 0$^{-+}$\ \  & $2.560\pm 0.155$\ \  &  $8.762\pm 1.089$
\\
&  & \ \ 2$^{-+}$\ \  & $3.040\pm 0.190$\ \  &   \\
&1D\ \ \ & \ \ 0$^{*++}$\ \  & $2.670\pm 0.310$\ \  &  $12.567\pm 1.702$
\\
&  & \ \ 3$^{++}$\ \  & $3.670\pm 0.230$\ \  &\ \   \\
\hline
$ggg$&1S\ \ \ & \ \ 1$^{--}$\ \  & $3.830\pm 0.230$\ \  &
$16.720\pm 1.967$ \\
&  & \ \ 3$^{--}$\ \ & $4.200\pm 0.245$\ \  &  \\
\end{tabular}
\end{ruledtabular}
\end{table}

\end{document}